# THERMODYNAMICS AND TRANSPORT PROPERTIES OF INTERACTING SYSTEMS WITH LOCALIZED ELECTRONS

A.L. EFROS

Department of Physics, University of Utah,
Salt Lake City, UT 84112 USA

## INTRODUCTION

The problem of strongly correlated electrons is a focus of modern condensed matter physics. Nobody doubts that the fractional quantum Hall effect results from electron-electron interaction. Many people thing that this interaction is also the origin of the high temperature superconductivity and the two-dimensional Insulator-Metal transition.

In this paper we concentrate on a problem of localized in space interacting electrons in a disordered system. The impurity bands of lightly doped semiconductors, the two-dimensional electron gas in different structures in an insulating regime are common examples of such systems.

The study of electron-electron interaction in localized regime has been initiated by M. Pollak [1] and G. Srinivasan [2]. Efros and Shklovskii [3],[4] have argued that the single particle density of states (DS) $G(\varepsilon)$ tends to zero at the Fermi level due to the long range part of the Coulomb interaction which, in a sense, remains non-screened. They proposed the following *universal* soft gap of the DS near the Fermi level at $T = 0$ which is called the Coulomb gap.

$$G(\tilde{\varepsilon}) = \frac{2}{\pi e^4}|\tilde{\varepsilon}| \text{ for } D = 2; \tag{1}$$

$$G(\tilde{\varepsilon}) = \frac{3}{\pi e^6}\tilde{\varepsilon}^2 \text{ for } D = 3. \tag{2}$$

The reference point for the energy $\tilde{\varepsilon}$ is the Fermi level. The *universality* means that the above DS is constructed from the energy $\varepsilon$ and from the electron charge $e$. Note that this is the only combination of $\varepsilon$ and $e$ with a proper dimensionality both in 2D and 3D cases. We assume here and below that the dielectric constant of a lattice is included into electron charge.

The lack of screening is connected with vanishing of the DS at the Fermi level. In this case the screening radius should depend on the magnitude of the electric field which is subjected to screening. Both the Coulomb gap and the non-linear screening result from the Coulomb law and from the discrete nature of electron charge. The question of screening in a system of localized interacting electrons has been considered in details by Baranovskii *et al* [5].

The main tool for a quantitative study of the Coulomb gap is the computer simulation [6-14], which mostly confirm the above results on the DS. However, some deviations from the universal behavior have been reported [9,13] and even the main concept has been equivocal [15]. We think that in 2D case these deviations can be understood in terms of the work [14], which shows that the universal physics of the Coulomb gap in the 2D case is valid at large disorder only. The apparent proximity to the universal result has been observed earlier because the simulations were restricted by a relatively high values of a disorder.

A renormalization group calculations, performed by Johnson and Khmelnitskii [16], confirmed Eqs.(1-2) for strong disorder and the results of Ref. [14] on the crossover from strong to weak disorder.

The experimental manifestations of the Coulomb gap are mostly variable range hopping conductivity (VRH) and tunneling experiments. Without interaction the VRH conductivity obeys the Mott law [17]

$$\sigma \sim \exp - \left(\frac{T_M}{T}\right)^s. \tag{3}$$

Here $s = 1/4, 1/3$ for three-dimensional (3D) and two-dimensional (2D) systems respectively, $T_M = \alpha_D/G_0 \xi^D$, where $G_0$ is the DS at the Fermi level, $\xi$ is localization length, $\alpha_D$ is a numerical coefficient which depends on the space dimensionality $D$.

It is obvious that Eq. (3) can not be valid if the Coulomb interaction makes $G_0 = 0$. In the next section we present the derivation by Efros and Shklovskii [3], which gives the law

$$\sigma \sim \exp - \left(\frac{T_0}{T}\right)^{1/2} \tag{4}$$

for $D = 2, 3$. Here $T_0 = \beta_D e^2/\xi$, where $\beta_D$ is another numerical coefficient which depends on the space dimensionality.

In some cases, like in neutron-transmutation-doped germanium [18], one can be sure that experimental data closely follow the Efros-Shklovskii (ES) law while in some other cases, where conductivity changes only by a few orders of magnitude, it is difficult to distinguish between the two laws. We think that the most important results have been obtained recently by groups of Jiang and Dahm [19] and Adkins [20]. They have proved that the transport they observed in 2D case reflects the crucial feature of the Coulomb gap, *the lack of screening in a system of localized electrons*. To do that they used a metallic electrode (gate) parallel to the 2D electron gas at a distance $d$ from the gas. This electrode provides an extra-screening of the Coulomb interaction. Namely, at a distance, larger than $d$, the interaction between electrons becomes dipole-dipole due to the image charges in the metal. Then, the Coulomb gap becomes smeared and the DS is energy independent at $|\varepsilon| < e^2/d$. Therefore, the conductivity obeys the ES law at high temperatures and the Mott law at low temperatures. Note, that without an external screening the Mott law may be valid at high temperatures where the relevant electronic states are outside the Coulomb gap while the ES law is valid at low temperatures. The experimental observation of the completely different and very unusual temperature dependence in the gated structures confirms that the the long-range interaction plays an important role in the VRH. The study of the $d$- dependence of the VRH gives even stronger arguments in favor of the Coulomb gap.

The direct observation of the Coulomb gap in tunneling has been claimed by Lee and Massey [21-22].

The glassy properties are another important manifestation of the electron-electron interaction in a system with localized electrons. Davies, Lee and Rice [8] where the first to rise this problem theoretically. They have coined a term "electron glass" which have survived until now, sometimes with a transformation to "Coulomb glass". These terms have to stress the relation of an electron system and a spin glass system. Davies *et al.* introduced an analog of the Edwards-Anderson order parameter, but their calculations already showed that this parameter is, probably, non-zero at any temperature. The absence of such parameter in the system with an external (or built-in) disorder has been pointed out by many researches (See [23] and references therein). However, the glass transition may exist even without Edwards-Anderson order parameter.

The new experiments, started by the group of Ovadyahu in 1993 [24,25], definitely show the relation of this system and the ordinary glasses. In these experiments one could control the electron density in the films of indium oxide by the gate voltage $V_g$. In some density range the mechanism of a conductivity is the VRH, which means that electron states are localized. The most important result is the discovery of a memory in this range. The sample, slowly cooled down at some $V_g = V_0$, memorizes the value of $V_0$. Conductance at low temperature, measured as a function of $V_g$, has a small minimum at $V_0$. The minimum disappears with time very slowly. The characteristic time is up to 15 hours and increases in a magnetic field. Similar phenomena have been observed by the group of A. Goldman [26] on ultrathin films of metals near the superconductor-insulator transition. Slow relaxation has been demonstrated by Don Monroe *et al.* [27] in compensated GaAs.

The analogy with the ordinary dielectric glasses can be seen from the data of Osheroff group [28], where similar effect in the capacitance of silicon oxide has been reported.

As the first step toward the understanding of the glassy effects we have studied the thermodynamic properties of the Coulomb glass[29].

The paper is organized as follows. In the second section we give a simple derivation of the Coulomb gap and of the VRH of interacting electrons. In the third section we concentrate on the glassy properties of an electron system.

## COULOMB GAP AND VRH CONDUCTIVITY

### Hamiltonian and Some Exact Properties

We consider here a standard classical Coulomb glass Hamiltonian [4]

$$H = \sum_i \phi_i n_i + \frac{1}{2} \sum_{i \neq j} \frac{e^2}{r_{ij}} (n_i - \nu)(n_j - \nu). \tag{5}$$

The electrons occupy sites on a lattice, $n_i = 0, 1$ are the occupation numbers of these sites and $r_{ij}$ is the distance between sites $i$ and $j$. The quenched random site energies $\phi_i$ are distributed uniformly within the interval $[-A, A]$. To make the system neutral each site has a positive background charge $\nu e$, where $\nu$ is the average occupation number, i.e. the filling factor of the lattice. If $\nu = 1/2$, the chemical potential $\mu$ is zero at all temperatures due to electron-hole symmetry.

Hereafter we take the lattice constant $l$ as our unit length and $e^2/l$ as our unit energy. Using these units, the single particle energy at site $i$ is given by

$$\varepsilon_i = \phi_i + \sum_j \frac{1}{r_{ij}} (n_j - \nu). \tag{6}$$

An important result for the Coulomb glass Hamiltonian is that the average occupation number of a site with energy $\varepsilon$ is given by the Fermi function. This result can be obtained from a self-consistent equation derived in Ref. [30]. It has been also mentioned as an exact result in Ref. [31]. Due to the strong electron-electron interaction this is not obvious, and we provide a short proof here, which has never been published before. The average occupation number of sites with energy $\varepsilon$ can be calculated by considering a single site $i = 1$ and calculating it's average occupation number $< n_1 >$ subject to the constraint that it has the required energy $\varepsilon$. By definition, this is given by

$$< n_1 > \;=\; \frac{\text{Tr}\,(n_1 \exp(-(H - \mu \sum_i n_i)/T)\delta(\varepsilon - \varepsilon_1))}{\text{Tr}\,(\exp(-(H - \mu \sum_i n_i)/T)\,\delta(\varepsilon - \varepsilon_1))}. \tag{7}$$

The Hamiltonian Eq. (5) can be written in the form $H = n_1 \varepsilon_1 + H'$, where $H'$ does not depend on $n_1$. This enables us to separate out $\text{Tr}_1$ which is the trace over the variable $n_1 = 0, 1$, thus obtaining

$$\begin{aligned}
< n_1 > \;&=\; \frac{\text{Tr}_1\,(n_1 \exp(-n_1(\varepsilon - \mu)/T))\,\text{Tr}'\,(\exp(-(H' - \mu \sum_i' n_i)/T)\delta(\varepsilon - \varepsilon_1))}{\text{Tr}_1\,(\exp(-n_1(\varepsilon - \mu)/T))\,\text{Tr}'\,(\exp(-(H' - \mu \sum_i' n_i)/T)\delta(\varepsilon - \varepsilon_1))} \\
&=\; \frac{\text{Tr}_1\,(n_1 \exp(-n_1(\varepsilon - \mu)/T))}{\text{Tr}_1\,(\exp(-n_1(\varepsilon - \mu)/T))}.
\end{aligned} \tag{8}$$

Here $\text{Tr}'$ and $\sum'$ stand for the trace and sum over all $n_i$ except $n_1$. From Eq. (8) we readily obtain the Fermi function

$$< n_1 > \;=\; f(\varepsilon) \equiv \frac{1}{1 + \exp(\varepsilon - \mu)/T}. \tag{9}$$

**Coulomb Gap in a Single-Particle Approximation**

The Coulomb gap in the DS around the Fermi level can be derived in a simple manner [32].

At zero temperature the distribution of electrons over the lattice sites is determined by the condition of minimum $H$ at a given total number of electrons. Equivalently, one can look for the unconditional minimum of the functional

$$\tilde{H} = H - \mu \sum_i n_i. \tag{10}$$

The single-particle energies $\tilde{\varepsilon}_i$ have the reference point at the Fermi level $\mu$

$$\tilde{\varepsilon}_i = \varepsilon - \mu. \tag{11}$$

The functional $\tilde{H}$ should be minimized with respect to simultaneous changes of any amount of occupation numbers $n_i$. It is easy to see that the change of one occupation number gives the condition $n_i = 0$ if $\tilde{\varepsilon}_i > 0$ and $n_i = 1$ if $\tilde{\varepsilon}_i < 0$, which is equivalent to the regular definition of the Fermi level $\mu$ in a non-interacting Fermi gas.

In the next approximation we consider the transfer of one electron from the site $i$, occupied in the ground state, to the site $j$, which is vacant in the ground state. The energy increment of $\tilde{H}$ is positive if for any pair of such sites

$$\omega_{i,j} = \tilde{\varepsilon}_j - \tilde{\varepsilon}_i - e^2/R_{i,j} > 0. \tag{12}$$

where $R_{ij} = \mathbf{r_i} - \mathbf{r_j}$. The last term in Eq. (12) reflects a simple fact that the ground state energy $\tilde{\varepsilon}_j$ includes the potential of electron, which is initially at site $i$.

One can see the origin of the Coulomb gap from Eq. (12). Since $\tilde{\varepsilon}_i < 0$ and $\tilde{\varepsilon}_j > 0$, the first two terms give a positive contribution, while the third term is negative. Thus, the sites with energy close to the Fermi level should be separated in space. Consider sites whose energies fall in a narrow band $(-\tilde{\varepsilon}/2, +\tilde{\varepsilon}/2)$ around the Fermi level. According to Eq. (12), any two sites in this band with energies on the opposite sides of the Fermi level must be separated by a distance $R_{ij}$ not less than $e^2/\tilde{\varepsilon}$. Therefore the concentration of such sites $n(\tilde{\varepsilon})$ cannot exceed $(\tilde{\varepsilon}/e^2)^D$, where space dimensionality $D = 2,3$. Then, the DS $G(\tilde{\varepsilon}) = dn/d\tilde{\varepsilon}$ must vanish at $\tilde{\varepsilon} \to 0$ at least as fast as $\tilde{\varepsilon}^{D-1}$.

It is clear that in the approximation based on Eq. (12), no faster law can arise than $\tilde{\varepsilon}^{D-1}$. Indeed, if it were the case, the interaction energy between the sites would be much less than their energies $\tilde{\varepsilon}$. Such a week interaction could not be responsible for lowering of the DS.

It is obvious from the above derivation that the long range Coulomb interaction is a crucial condition for the Coulomb gap. The DS at the energy $\tilde{\varepsilon}$ is determined by interaction at a distance $e^2/\tilde{\varepsilon}$. It has been show by computations that if the localized electrons are embedded into some conducting medium which is able to screen interaction, the Coulomb gap disappears [10].

This consideration leads to Eqs. (1,2), where numerical coefficients are obtained from the mean-field equation [4,10]. The Eqs. (1,2) are valid if $G(\tilde{\varepsilon}) \ll G_0$, where $G_0$ is a bare DS without interaction. In our model it is $1/Al^D$, where $l$ is the lattice constant. Thus, at large $A$ the width of the Coulomb gap is $E_g = (e^2/l)^D (1/A)^{D-1}$. At $|\tilde{\varepsilon}| \gg E_g$ the DS $G(\tilde{\varepsilon})$ is close to $G_0$.

The applicability of the above consideration is limited by the condition $A \gg e^2/l$ or in dimensionless units $A \gg 1$. The crossover between large and small disorder has been considered in Ref. [14] for $D = 2$. It has been shown that $A = 1$ can be considered as a large $A$, since substantial deviation from the law Eq. (1) appears at very low energy only.

The dimensionality $D = 1$ is a marginal for the Coulomb gap. In this case one gets [33]

$$G(\tilde{\varepsilon}) = \frac{G_0}{1 + G_0 e^2 \ln(\frac{e^2}{l\tilde{\varepsilon}})}. \tag{13}$$

Thus, DS tends to zero at $\tilde{\varepsilon} \to 0$ as $1/\ln(1/\tilde{\varepsilon})$. This result is a starting point for an $\varepsilon$-expansion used in Ref. [16].

At finite temperature, it has been shown [8,10] that the gap is smeared at energies smaller than the temperature. Roughly speaking, $G(\tilde{\varepsilon}) \sim G(T)$ if $T \gg |\tilde{\varepsilon}|$.

**Interaction of Excitations**

Until now we have taken into account only conditions of the minimum of $\tilde{H}$ with respect to change of one and two occupation numbers. What about many electron transitions? Do they present an extra restrictions on the DS or everything is already taken into account? The analysis of this questions can be done in terms of interaction of dipole and charge excitations. It shows that in the 2D case the single-electron approach, used above, is, probably, good, while in the 3D case physics is more difficult.

All single-electron excitations are electron transitions from the sites, which are occupied in the the ground state, to the empty sites. The energy of the transition of one electron is given by Eq. (12).

For any transition from the ground state the energy $\omega_{i,j}$ should be positive. The result of each transition is an "electron-hole pair". If $\omega \gg e^2/R$, the pair is just two independent single-particle excitations: an extra electron on a site $j$ and an extra hole on a site $i$. They participate in the VRH as the carriers and they are described by the single-particle DS.

If $\omega \ll e^2/R$, the pair is rather an exciton with a large binding energy. From electrostatic point of view this is a small dipole. A dipole can be also formed by transitions of many electrons in a small region. Such a soft excitation has many-electron nature. At small $\omega$ these dipoles are spatially separated from each other and they do not participate in dc. However, they contribute to the ac and they are responsible for the low temperature thermodynamics of the system. The interaction of these dipoles has been carefully studied (See review [10]) and has been found not very important even in the 3D case. Assuming that the DS of the dipoles is a weak function of energy, one gets that the specific heat provided by these excitations should be approximately linear in temperature. The typical size of a pair is $r_0 = e^2/E_g$.

The crucial question now is the interaction of the single particle excitations (carriers) with this dipoles. Suppose, an extra electron enters the system or a pair with a very large $R$ is excited. In both cases an extra electron appears at some point which we take as the origin. This electron creates an electric field $E = e/r^2$. The field causes electron transitions leading to a new ground state. At a large distance the field is small and transitions occur within the pairs with a small excitation energy. The transition occurs if $e\mathbf{E} \cdot \mathbf{R} > \omega$. Here $e\mathbf{R}$ is the dipole moment of the pair with the excitation energy $\omega$. As a result, a pair becomes polarized by the electric field. It is easy to show that the total number of dipole pairs polarized by this field at a distance $r$ is of the order of $(r/r_0)^{(D-2)}$, where $D$ is space dimensionality. In these estimates we use for the DS of pairs $g = G_0$ and we ignore the logarithmic factor in the DS, which may be due to the interaction of pairs in 3D case [10].

At $D = 2$ the interaction between carriers and dipoles is not very important. We think, however, that it creates a persistent drift between the pseudoground states observed recently [29] (See the next section).

However, at $D = 3$ the number of polarized pairs is large and they create a polaronic shift of the order of the width of the Coulomb gap $E_g$ [10]. That is why in 1980 we came to conclusion [34] that carriers have very different nature in 3D- and 2D-cases and that in 3D-case they are rather "electronic polarons" than individual electrons. It follows straightforward [10] that for $D = 3$ the single-particle DS has a form $G \sim \exp(-E_g/|\tilde{\varepsilon}|)$ instead of Eq. (2).

It is obvious now that the concept of "electron polarons" does not correspond to reality. The most accurate computational results [9] are inconsistent with exponential behavior of the DS in 3D-case. Experimental data also do not show any signature of the polaronic effect. Thus, we have a contradiction in our understanding of the 3D-case.

We would like to propose a new picture of the low energy excitations in the 3D-case [35]. The basis of this approach is a generalized universality principle. Assume that at large disorder the DS of pairs with given energy and length is universal. We are considering now the DS of all low energy excitations at a given length $R$, in the interval of $\omega$, which are either carriers or dipoles. In some way this description takes into account interaction between pairs of all sizes, and the proposed DS is a result of this interaction. In 3D-case only the expression

$$g(\varepsilon, R) = \alpha \varepsilon^q (\varepsilon + e^2/R)^{2-q}/e^6 \qquad (14)$$

is universal and obeys the proper dimensionality(1/energy· volume). Universality means that Eq. (14) contains only the charge of electron. Here $\alpha$ and $q$ are some numerical constants, and $\varepsilon$ is a positive energy of an excitation. Eq. (14) is valid if both $\varepsilon$ and $e^2/R$ are less than $E_g$. At large $R$ we get parabolic DS for carriers, which coincide with Eq. (2). At small $R$ one gets for the DS of short pairs $g = \alpha \varepsilon^q (e^2/R)^{(2-q)}/e^6$. Then, specific heat $C_V/T \sim T^q$.

To choose $q$ we should assume that the polaronic effect is absent. Otherwise Eq. (14) would be inconsistent because of the polaronic shift in a single-particle DS.

It follows from Eq.(14) that the total number of pairs $N_p$ with the length $R$ polarized by one extra electron at a distance smaller than $r$ is

$$N_p \approx g(e\mathbf{E} \cdot \mathbf{R}, R)(e\mathbf{E} \cdot \mathbf{R})r^3 \sim (r/R)^{1-2q}. \qquad (15)$$

The concept of constant DS for small dipole pairs and of a strong polaronic effect corresponds to $q = 0$. For the new scenario we choose $q = 1/2$. Then the dipoles are not interacting with electrons in all scales or, to be precise, $q = 1/2$ is a marginal exponent for the interaction. In this sense the picture is the same as in 2D case. One should mention at this point that we try to avoid the linear divergence in $N_p$ at large distances and, on this stage, we do not care about possible logarithmic factors.

Finally, the new scenario for 3D-case gives that the DS of short pairs $g = \alpha \varepsilon^{1/2}/R^{3/2}e^3$. The pairs with $R = r_0 = e^2/E_g$ give the major contribution to thermodynamics. The DS of carriers obeys Eq. (2).

An appealing feature of the proposed scenario is the unification of all strong interactions. Consider all dipoles and carriers with the energy less than $\varepsilon$. The distance between the dipoles is of order of $r_p = R^{1/2}e/\varepsilon^{1/2}$. At small $\varepsilon$ this distance is much less than the distance between carriers $r_c = e^2/\varepsilon$. However, the interaction between the dipole with length $R$ and the nearest carrier is of the order of $e^2 R/r_p^2 = \varepsilon$, which is of the same order as interaction between the nearest carriers. Thus, all important interactions become of the same order.

If this scenario is correct, one should expect that specific heat $C_V \sim T^{3/2}$. It might be possible to check it experimentally in lightly doped semiconductors. It is possible also to check it by a standard computer modeling at finite $T$. Such a modeling has been done before by different methods [23,36-37]. The results show superlinear temperature dependence. Möbius and Pollak [37] got $C_V \sim T^p$ with $p \approx 1.8 \pm 0.2$. This is close to the result we expect.

The main problem for the modeling is a large size effect due to the long-range interaction. As far as we know, the size effect in the specific heat has never been studied systematically. It can be described in the framework of the above scenario. The macroscopic regime appears when the size of the system is much larger than the average distance between the pairs $r_p = r_0^{1/2}e/T^{1/2} = e^2/\sqrt{E_g T}$, We predict that in a sample $L \times L \times L$

$$\frac{C_V}{T} = \frac{T^{1/2}}{e^3 r_0^{3/2}} f\left(\frac{r_0^{1/2}e}{T^{1/2}L}\right), \tag{16}$$

where $f(x)$ is some constant at $x \ll 1$ and $f(x) \sim x$ at $x \gg 1$. The scaling law, given by Eq. (16), is the best check of the proposed scenario. For this purpose one should neither go very low in temperature no increase the size very strongly. One should just show that $C_V/T$ is a universal function of $T^{1/2}L$. In the same finite temperature computer modeling one can find directly the form of the function $g(\varepsilon, R)$.

An experimental evidence for this scenario may come from the measurements of the ac conductivity of disordered systems. The theory of ac conductivity goes back to the famous papers by Pollak and Geballe [38] and Austin and Mott [39]. It has been reconstructed by Efros and Shklovskii [10] taking into account interaction. However, the constant DS of short pairs has been assumed. If the $q = 1/2$-scenario will be proved, all the theory should be reconstructed again. This reconstruction will substantially change the temperature and frequency dependencies of the conductivity.

We understand that the Nature might be more sophisticated than any scenario. Say, another possibility for the 3D case would be a glass transition at finite temperature. Finally, we think that the problem of the DS in 3D case is very important and still unsolved.

**Variable Range Hopping Conductivity (VRH)**

In 1968 N. F. Mott [17] found out that at sufficiently low temperatures hopping conduction results from the states, whose energies are in a narrow band around the Fermi level. With decreasing temperature the width of the band decreases and the hopping length increases. That is why this mechanism is called VRH.

Mott assumed that the DS in the relevant band near the Fermi level is energy independent. To repeat his calculations assume that the hops are allowed within the band $(\mu + w, \mu - w)$ only. The concentration of sites within the band is $N_w = 2wG_0$ and the average distance between them is $R \sim N_w^{-1/D}$. Hopping rate has a form

$$W = W_0 \exp\left(-\frac{2R}{\xi} - \frac{w}{T}\right), \tag{17}$$

where $R$ is the distance between the two sites and $\xi$ is the localization length. The first term in the exponent is responsible for the tunneling, while the second one describes activation. We do not care now about numerical factors which cannot be obtained by this way.

To estimate the logarithm of effective conductivity $\sigma_w$ due to the hops inside the chosen band one may substitute an average distance between the sites instead of $R$. Then

$$\ln \sigma_w \sim -\frac{1}{\xi(wG_0)^{1/D}} - \frac{w}{T}. \tag{18}$$

The tunneling term is small at large $w$, while activation prefers small $w$. The exponent has a maximum at $w_e = T^{3/4}/(G_0\xi^3)^{1/4}$. This is the width of the VRH band. At this value of $w$ one gets the Mott law Eq. (3). To calculate numerical coefficient $\alpha_D$ one should solve the corresponding percolation problem [40].

To take into account the Coulomb gap near the Fermi level one should use

$$N_w = \int_{-w}^{w} G(\tilde{\varepsilon}) d\tilde{\varepsilon}. \tag{19}$$

Repeating similar calculations, one gets Eq. (4). Note that in this case $w_e \approx \sqrt{T_0 T} \gg T$. Therefore we may not to take into account the smearing of the Coulomb gap due to the finite temperature.

The first principal computer simulation of the VRH in the interacting system is an extremely difficult problem because of a strong size effect and a huge dispersion of transition rates given by Eq. (17). We think that all attempts to solve this problem, which we know, [7,41,42] are not satisfactory due to different reasons. In Ref. [7] we artificially slow down the transition rates of small pairs, while in Ref.[41,42] the size of a system is too small. The experimental situation is described in the Introduction.

## THERMODYNAMIC FLUCTUATIONS OF SITE ENERGIES AND OCCUPATION NUMBERS

In this section we are trying to understand the nature of a glassy properties observed experimentally in Ref. [24-27]. This part is based upon our works [29,43].

Another important effect, coming from the interaction, is that the phase space of the many particle system has the so-called pseudoground states (PS) which were first described by Baranovskii *et al.* [6] and then studied by many authors [44-46] in connection with the long range relaxation.

The PS are the states with the total energies very close to one another but with very different sets of the occupation numbers. Thus, many electrons have to be transferred to go from one state to another. Each state presents a local minimum of the total energy and transition from one state to another requires passing huge barriers, if electrons are moving by small groups one after another.

We understand the slow relaxation of a glassy system as traveling around different PS which is hindered by barriers between them. We study the system of interacting electrons in thermal equilibrium, where it has a possibility (long enough time) to wander around many PS. Such a wandering is accompanied by fluctuation of site occupation numbers. The main manifestation of these fluctuations is that the configuration of occupied sites within the Coulomb gap changes with time, even though the shape of the gap itself is time independent. This persistent change of the configuration of occupied sites occurs even at temperatures which are much lower then the Coulomb gap width. A related effect is the fluctuations in the site energies, the magnitude of which is much larger than the temperature.

A crucial ingredient of the above picture is the assumption that there is no finite temperature thermodynamic glass transition in the system. In the thermodynamic limit, such a transition would prevent the system from reaching thermal equilibrium below the transition temperature, thus limiting the validity of our results to finite sized samples. In fact, no such transition has been observed either experimentally or numerically in the two dimensional (2D) Coulomb glass. Furthermore, our system has much in common with various 2D spin glass models, where there is a strong numerical evidence that no finite temperature thermodynamic transition occurs [47]. In the work [43] we present results which support such a conclusion for the Coulomb 2D glass as well,

While we believe that the 2D Coulomb glass can always reach thermal equilibrium, the experimentally observed glassy dynamics [24-25] indicate that at low temperatures the equilibration time of the system becomes very long. This results from the large energy barriers that need to be traversed in order for the system to drift amongst the different PS's. As the aim of the current work is to study thermodynamic fluctuations, it is necessary to design the simulations so that the system equilibrates within a computationally feasible time scale. Since the standard Coulomb glass Hamiltonian itself does not contain any dynamics, we employ dynamics which differ from the physical dynamics of a typical system, but are significantly faster. Namely, we assume that the transition rate is independent of a distance between sites. Therefore, any non-equilibrium phenomena, and particularly transport, cannot be studied directly by the methods discussed here.

**Spectral Diffusion**

Now we turn to the central topic of this Section, which is the study of the thermodynamic fluctuations within the Coulomb gap. These fluctuations can be seen in a few ways. One is the time dependence of the single particle energies, which we call spectral diffusion.

Our computer simulations use the standard Metropolis algorithm, where the rate of a hopping transition depends only on the energy difference between initial and final configurations. The simulations were performed on a square lattice of $L \times L$ sites with periodic boundary conditions. In this torus geometry, the distance between any two sites is taken as the length of the shortest path between them. The filling factor $\nu = 1/2$. All results obtained were averaged over $P$ different sets of the quenched random energies $\{\phi_i\}$. Unless stated otherwise the value $P = 100$ was used throughout. Furthermore, it was verified that all our results saturate as a function of the system size.

To study the spectral diffusion, we first equilibrate the system for $t_w$ MC sweeps. (We define a single MC sweep as a series of $N = L^2/2$ consecutive MC attempts). Then we mark all the sites whose single particle energies are in a narrow interval $[E_c - W, E_c + W]$ within the Coulomb gap as "test sites". We follow the evolution of the distribution of these energies as the simulation proceeds. We observe that after some number of MC steps, this distribution becomes time-independent. We have also checked that the final form of the distribution does not depend on the initial time $t_w$.

This asymptotic form is shown in Fig. 1 for $E_c = 0$, $W = 0.1$, at two different temperatures. Note that the final energy distribution covers most of the Coulomb gap, although the initial distribution (shown by arrows in Fig. 1) is centered in a small region at the center of the gap. This is in spite of the fact that the temperature is much smaller than the gap width. We have also studied the dependence of the final distribution on $W$, and have found that the results are independent of $W$ for $W < 0.1$. Also shown in Fig.1 is an example where the initial distribution is asymmetric, namely $E_c \neq 0$. In this case, the asymptotic distribution is also asymmetric, however it is nearly as broad as the symmetric distributions. Moreover, while the initial asymmetric distribution consists of only sites with positive energies (unoccupied sites), the final distribution also includes sites with negative energies (occupied sites).

Another way to observe spectral diffusion is to measure the time average of the single-particle energy at site $i$, $\langle \varepsilon_i \rangle$, and the standard deviation at the same site, $\Delta_i = \sqrt{\langle \varepsilon_i^2 \rangle - \langle \varepsilon_i \rangle^2}$. We perform this calculation for all sites and create a function $\Delta(\langle \varepsilon \rangle)$. This function is shown in Fig. 2 for $A = 1$ and several temperatures. It is found that for all sites, the standard deviation of the single-particle energies is much larger than the temperature. Moreover, for sites with energies near the Fermi level the standard deviation is up to two times larger than for other sites.

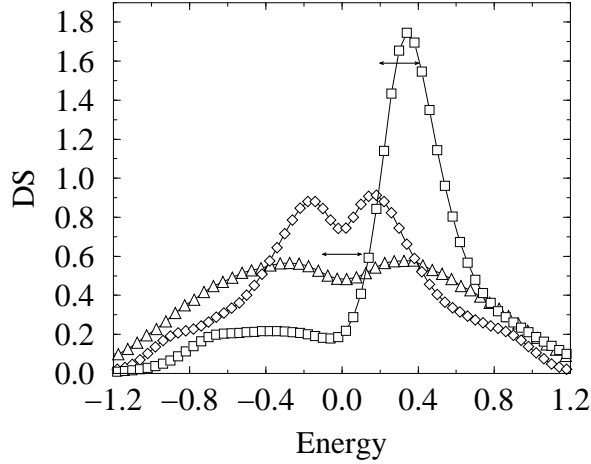

**Figure 1.** Final energy distribution of sites initially in the energy range $[E_c - W, E_c + W]$. Diamonds are for $E_c = 0, W = 0.1, T = 0.05$. Squares are for $E_c = 0.3, W = 0.1, T = 0.05$. Triangles ($\triangle$) are for $T = 0.1, E_c = 0$ and $W = 0.1$. The arrows mark the positions of the two initial distributions of test sites which were used. All results are for $A = 1$.

We understand this picture in the following way. Sites with large $\Delta$ are expected to be "active" sites which change their occupation frequently, as their energies cross the Fermi level. The changes of the occupation numbers of these sites are accompanied by a reorganization of the local configuration of occupied sites, which in turn is responsible for the larger value of $\Delta$ in a polaron manner. On the other hand, the sites with smaller $\Delta$ are "passive" sites, and they change their energy only in response to the random time dependent potential created by the active sites. In the context of passive and active sites, it is instructive to define the quantity $E_w$ as the energy at which $\Delta(\langle \varepsilon \rangle) = \langle \varepsilon \rangle$. The meaning of this is that sites which satisfy $\langle \varepsilon \rangle < E_w$ have energy fluctuations larger then their average energy, and therefore are active. From Fig. 2 it is also apparent that these sites have larger value of $\Delta$, thus supporting our understanding that these are indeed the active sites of the system.

The width of the maximum in Fig. 2 may indicate that the active sites are predominantly within the Coulomb gap. This is reasonable, since the occupation number of sites within the gap is strongly affected by interactions. However, at $A = 1$ all characteristic energies, including the gap width, are of the same order. To check whether the active sites are indeed within the gap, we estimate the dependence of $\Delta$ on $A$ for $A > 1$ and compare it with simulations. The width of the gap $E_g$ decreases with $A$ as $E_g \sim 1/A$. The electron density within the gap is $n_g \sim \int_0^{1/A} \varepsilon d\varepsilon \sim 1/A^2$. If active sites make up a finite portion of all sites in the gap, they create time dependent potential with the mean square value $\Delta^2 \sim n_g \int_1^A r^{-2} r dr \sim \ln A / A^2$. We cut off the logarithmic integral at the screening radius which is proportional to the reciprocal density of states $A$. We make simulations for $1 \leq A \leq 4$ and plot the results of simulations in the scale $\Delta A / \sqrt{b \ln A + 1}$ against $\langle \varepsilon \rangle A$. The temperature for each value of $A$ is $T = 0.05/A$, keeping it constant in units of the gap width. If our hypothesis is correct we can chose parameter $b$ in such a way that all curves collapse in one at least for passive sites. One can see

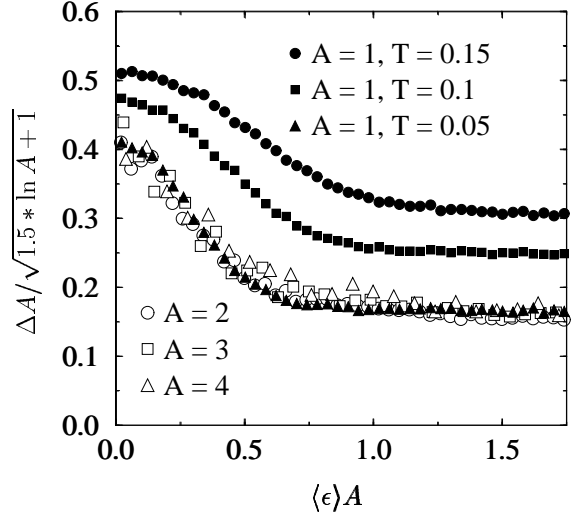

**Figure 2.** Site energy standard deviation as a function of site average energy $\Delta(\langle\varepsilon\rangle)$, for various values of $A$ and $T$. For $A > 1$ the temperature is given by $T = 0.05/A$. Only positive energies are shown due to particle hole symmetry.

from Fig.2 that indeed all curves collapse in one for $b = 1.5$.

The temperature dependence of the results of Fig. 2 is presented in Fig. 3. The curves marked $\Delta_{max}$ and $\Delta_{min}$ show the maximal and minimal values of the standard deviation $\Delta(\langle\varepsilon\rangle)$ as a function of temperature. The curve marked $E_w$ shows the width of the function $\Delta(\langle\varepsilon\rangle)$, which was previously defined. All the quantities shown in Fig. 3 appear to be much larger than the temperature for the entire temperature range shown.

A large number of active sites we understand as a result of wandering of the system around different PS. Each PS is characterized by a unique set of sites forming the Coulomb gap. So when the system passes from one PS to another the number of sites that change their occupancy is a finite fraction of the number of sites in the Coulomb gap. The energy separation between different PS decreases with increase of the sample size [43]. If the thermodynamic fluctuations of the total energy $\sqrt{C T}$ ($C$ is the heat capacitance of the system) is larger than this energy separation, then the described mechanism of the spectral diffusion gives a temperature independent contribution to $\Delta$ estimated above. For small samples, that we use for simulations, $\Delta$ goes to zero with temperature when $\sqrt{C T}$ becomes smaller than the energy separation between PS, and the above estimate does not work at too low temperatures. To get a non-zero result at the limit $T \to 0$ one should increase the size of the sample while decreasing temperature. We are not able to perform this procedure in full scale but we check that there is no size effect in the temperature range presented in Fig. 3.

We think that the temperature dependence in Fig. 3 results from soft dipole excitations (See previous section). The density of states of the dipole excitations is $1/A$, so that the concentration of active excitations is $\sim T/A$. These excitations not only contribute themselves to the spectral diffusion but also induce an energy change of passive sites leading to linear temperature dependence of $\Delta_{min}$ in Fig. 3. Indeed, each dipole creates the potential $r_0/r^2$ at a passive site, where $r$ is the distance between the site and the nearest dipole and $r_0 \sim 1/E_g \sim A$ is the size of the dipole [10]. This estimate leads to a slope of $\Delta_{min}(T)$ dependence close to that in Fig. 3.

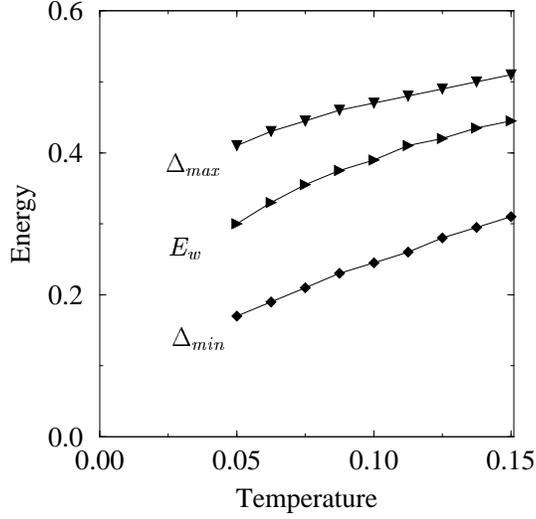

**Figure 3.** Temperature dependence of quantities $\Delta_{max}$, $\Delta_{min}$ and $E_w$ for $A=1$. See text for a description of these quantities.

## Correlation Function

Thus, the spectral diffusion shows that the configuration of occupied sites within the Coulomb gap persistently changes in thermodynamic equilibrium. To obtain more information about this motion, one can study the correlation function of occupation numbers. We do this by constructing a vector $\mathbf{D}(t_w)$ after $t_w$ MC steps have been performed. The components of $\mathbf{D}(t_w)$ are the occupation numbers $n_i$ of all sites within a given energy range $[-W,W]$. The vector is normalized so that $\mathbf{D}(t_w) \cdot \mathbf{D}(t_w) = 1$. As the simulation proceeds, we check the occupation number of these same sites, construct the vector $\mathbf{D}(t_w+t)$, and calculate the correlation function $C(t_w,t) = \mathbf{D}(t_w) \cdot \mathbf{D}(t_w+t)$. Correlation functions analogous to $C(t_w,t)$ are commonly used to measure the similarity between different configurations in systems such as spin glasses [47]. For two identical configurations $C(t_w,t) = 1$, while if there is no correlation $C(t_w,t) = 0.5$. Basically, we are interested in $C_\infty = \lim_{t_w \to \infty} \lim_{t \to \infty} C(t_w,t)$, which is a measure of the similarity of two arbitrary states of the system at thermal equilibrium. For a non-interacting system,

$$C_\infty = \frac{\int_{-W}^{W} f^2(\phi) d\phi}{\int_{-W}^{W} f(\phi) d\phi}, \qquad (20)$$

where $f(\phi)$ is the Fermi function. Thus, for the non-interacting system $C_\infty = 1 - T/W$ at $W \gg T$ and $C_\infty = 0.5$ at $T \gg W$.

In order to evaluate $C_\infty$ from the simulation, we measure $C(t_w,t)$ as a function of $t$ for a given $t_w$, and wait long enough so that $C(t_w,t)$ becomes independent of $t$. We denote this saturated value as $C(t_w,\infty)$. In the inset of Fig. 4 we show an example of such a saturation. We then increase $t_w$ until $C(t_w,\infty)$ becomes independent of $t_w$, and thus obtain our estimate of $C_\infty$.

The results for $C_\infty$ for the interacting system are shown in the main part of Fig. 4 as a function of temperature, for $A=1$, $W=0.3$ and $W=0.6$. The corresponding functions for the non-interacting system, calculated directly from Eq. (20), are also shown. We observe that the correlation for the interacting system is much weaker than for the corresponding

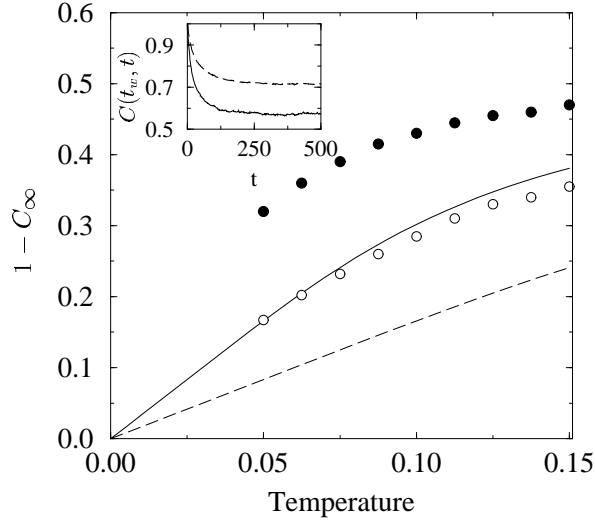

**Figure 4.** The correlation function $1 - C_\infty$ as a function of temperature for $A = 1$ and $W = 0.3$ (solid circles) and $W = 0.6$ (open circles). The solid line and dashed line show the correlation function for the non-interacting system, as given by Eq. (20), for $W = 0.3$ and $W = 0.6$ respectively. The inset demonstrates the saturation of the correlation function with time.

non-interacting system at the same temperature.

Note that by increasing $W$ we include more sites in the correlation function $C_\infty$. However, our results for the spectral diffusion indicate that most of these additional sites remain passive as $T \to 0$. Thus, $C_\infty$ should increase as $W$ increases, as we indeed observe comparing the results for $W = 0.6$ and $W = 0.3$ in Fig. 4.

The most interesting result coming from the study of the correlation function $C_\infty$ is the possibility of a finite value for $\lim_{T \to 0}(1 - C_\infty)$. Clearly, such an extrapolation cannot be considered conclusive, however, one has to keep in mind the following points: First, while we have included in the definition of the correlation function only sites that were in the initial energy range $[-0.3, 0.3]$, it is clear that many passive sites are still included in this definition. These passive sites mask the behavior of the active sites and tend to increase the correlation.

Second, a finite value of $\lim_{T \to 0}(1 - C_\infty)$ means that thermal motion continues down to zero temperature. This conclusion is consistent with the results obtained from the spectral diffusion in the previous section, and we understand it in the same way: Namely, we expect that a nonzero value of $1 - C_\infty$ may be obtained only if the size of the sample grows with the decrease of temperature so that $\sqrt{C}T$ remains larger than the energy separation between different PS.

It is important to point out that our results cannot be explained by assuming that the excitations of the system are separated pairs of sites, with electrons hopping back and forth between the sites of each pair. This assumption would mean that electrons are effectively localized in space. Since the energy density of such excitations is constant at low energies, meaning the number of available excitations decreases linearly with temperature, one immediately obtains that $\lim_{T \to 0}(1 - C_\infty) \approx T/W$, like in the non-interacting system. The same temperature dependence is obtained even if excitations involve a few electrons that change their positions simultaneously (so called many electron excitation [49-50]). In fact, any picture based upon confined separated excitations which do not interact with each other would mean that $\lim_{T \to 0}(1 - C_\infty) \approx T/W$. Since our data definitely contradicts this temperature dependence, we conclude that such excitations cannot explain our results.

Thus, a nonzero value of $\lim_{T\to 0}(1 - C_\infty)$ can be explained only by a thermal motion of such electron configurations in the whole system that cannot be separated into thermal motion of independent clusters. This means that there is a finite portion electrons that are not localized in some region of space but move around the whole system. We view this motion as another manifestation of wandering of the system around different PS.

The conclusions of this section are as follows. We have presented strong computational evidence that in a disordered 2D system of localized interacting electrons, the configuration of occupied sites within the Coulomb gap persistently changes with time. This effect persists down to temperatures well below the Coulomb gap width, and it causes a large time dependent random potential. This result is an exclusive property of the interacting system. Without interaction only electrons in a small interval of the order of $T$ change their occupation.

We argue that this effect may exist at zero temperature, if the size of the sample increases with decreasing temperature so that the separation between local minima of the total energy is much smaller than the thermal fluctuations of the total energy. However, our computational abilities are not enough to prove this point. Nevertheless, our simulations have been done at low enough temperature to compare them with the experimental data.

We have obtained these results for equilibrium state only. Our program does not limit the length of the electron transition in order to reach equilibrium by the easiest way. In a real systems hopping distance is limited by tunneling so that applicability of the above results to a real system arises questions.

We have shown [43] that equilibration time saturates at some value $L = R_c$, which is approximately proportional $1/T$. One can draw a conclusion that equilibration time is size-independent if localization length $\xi$ is large. Such a condition can be satisfied at very low $T$ near metal-insulator transition. However we are not able to discuss transport in this situation because our computations still do not take into account such important factors as narrowing of the Coulomb gap with increasing $\xi$ (See [22]) and decrease of the random potential because of the quantum mechanical averaging at large $\xi$.

## CONCLUSIONS

We have given a critical review of the theory and experiment on the interaction of localized electrons and hopping conductivity. There is another very interesting field of quantum mechanical description of the same system which includes such important problems as the metal-insulator transition. The simplest Hamiltonian of this problem can be obtained by adding the non-diagonal term $\sum_{i\neq j} J_{ij} a_i^\dagger a_j$ to the classical Hamiltonian Eq. (5). Many efforts have been made in this field, but we do not review them in this paper. We would mention only that quantum theory should include all the problems of classical theory mentioned above... and something else.

Coming back to the classical theory of the Coulomb gap, we would make the following conclusions:

1. We believe that the theory of the 2D systems is basically complete and correct. We are not so optimistic about the 3D systems. We believe that the physics of long-range interaction is valid in this case as well, we believe that there is a gap near the Fermi level, we even think that the VRH law Eq. (4) is also valid, but we are not sure about the structure of elementary excitations in this case and about the possibility of thermodynamic glass transition. However, we propose in this paper a reasonable scenario which leads to a pretty simple picture, similar to the 2D case.

2. We think that the pseudoground states are responsible for the glassy effects which has been observed recently. We claim that in 2D case there is no thermodynamic glass transition, but the equilibration time strongly increases at low temperatures. We argue that the VRH in the interacting system might be a result of non-ergodic behavior due to the long-time relaxation at low temperatures.

## AKNOWLEDGEMENTS

I am grateful to my long-term friend and coauthor Boris Shklovskii for many long discussions of the problems, considered here. I am grateful to David Menashe, Boris Laikhtman, and Ofer Biham, with whom I was working on the problem of thermodynamic fluctuations. I appreciate helpful discussions with Zvi Ovadyahu and A. Vakhin. The work was funded by BSF Grant 9800097.